# Co-creation of AI technology, empowering curators of cultural heritage information and guarding research commons

*Andrea Scharnhorst, Han Yang, Jetze Touber, Kim Ferguson, Philipp Mayr, and Vyacheslav Tykhonov*

## 1. Introduction – the use of large language models in engineering explorations around long-term archiving

The substance of this paper is the description of the use of Retrieval-Augmented Generation (RAG) – a new methodology for information retrieval – for specific digital collections of cultural assets. The collections are provided by institutions operating in the cultural sector. The topical areas are the humanities and social sciences. More concretely, most of the work presented here was enabled by a European-funded research project MuseIT[1] which is clearly situated in the realm of fostering new technologies for Cultural Heritage.

The International workshop on "Large Language Models for the History, Philosophy, and Sociology of Science" held in Berlin in April 2025[2] enabled us to present engineering explorations around implementing machine-aided knowledge organisation and local generative AI (chatbots). The presentation of our basically practice-born approaches in such a scientific context had various functions. We were able to see if our engineering explorations stand the test of a scientific debate and how they compared to other experimentations around LLMs. But, the setting of the workshop in the field of History, Philosophy, and Sociology of Science came naturally with a reflexive layer on the nature and role of LLMs for academia and beyond.

We adhere to this interaction by presenting a sequence of our experimentations. This sequence is narrated as a specific journey of engineering all executed around a spe-

1 Multisensory, User-centred, Shared cultural Experiences through Interactive Technologies (MuseIT) Project funded 2022 – 2025. Available at https://www.muse-it.eu/ (accessed November 5, 2025).

2 Programme of the workshop. Available at https://www.tu.berlin/hps-mod-sci/llms-for-hpss/workshop-llms-for-hpss (accessed November 5, 2025).

cific data-sharing and archiving platform *Dataverse*[3] (Crosas, 2017). Implementing a local chatbot for collections – a method also known as RAG in Information Retrieval, is the current culmination of this journey. To connect the engineering explorations with the scientific discourse and also with other contributions to these proceedings, we present at the end of this introduction a literature study on the RAG discourse.

In the next section of this paper, we point out how these RAG applications are actually the result of a longer process of enabling agile infrastructures for data sharing and archiving which feed on forefront technological developments, while at the same time adhering to principles of Open Science and empowering institutions in the public sector (Tykhonov, 2020). We return to this more general aspect in the conclusion section of this paper.

The engineering journey we describe in the core of the paper starts from 'archives for everyone' and ends with 'local chatbots for specific collections'. It has been executed with the help of various projects which all came with specific focal points. But, to some extent in one of the last projects in that row, MuseIT, all those earlier experiences came together, and so we use the 'MuseIT – case' to illustrate this engineering quest. The question underlying our own explorative journey has been always: How to improve the interaction with collections of documents – preferably archived in a way which respects the principles of FAIRness (Wilkinson et al., 2016) and Open Science (Bartling and Friesike, 2014). We use the *MuseIT lens* to unfold the various techniques which can be applied to increase accessibility of collections, from machine supported Knowledge Organization Systems (KOS) construction, connecting KOS services, to building search interfaces using a combination of LLM and KOS techniques. One could say this part of the paper is a 'practice description'. We use the concluding section of the paper to unfold a couple of more generic aspects in the description of our use-case.

With this concluding section we respond to the invitation which the editors of these proceedings formulated as such "Rather than offering a conventional edited volume or a delayed 'state of the art' account—likely outdated upon publication—it gathers timely, authentic contributions from a field in motion. This framing invites openness, experimentation, and personal interpretation, encouraging authors to draw on their own styles, interests, and positions within or adjacent to the HPSS community. It reflects the spirit of the workshop—lively, collaborative, and intellectually diverse—and carries that spirit into the published volume."[4] Taking this invitation to heart and still inspired by the workshop discussions themselves, we felt encouraged to add a more reflexive layer to our 'practice report'. We are not sure which of the possible writing-forms the editors listed as alternatives, "traditional essays, dialogues, creative thought experiments, research vignettes, theoretical provocations, agenda-setting pieces, tutorial-style texts, or autotheoretical pieces," best describes our reflections. In any case, we would like to share them in the context of these proceedings because they touch upon dimensions which might also have a relevance for other more methodological or theoretical contributions in these proceedings.

3 More information on Dataverse as a project, its use and best practices, the community around it, and the software. Available at https://dataverse.org/ (accessed November 5, 2025).

4 Personal communication, invitation to submit papers.

All of these additional dimensions, epistemologically spoken, concern science-of-science aspects. They are questions about the nature of knowledge production as executed in our experiments, the role of the stakeholders involved, and the quest for a terminology to discuss innovation in research infrastructures. They partly touch on old fundamental questions of the roles of intelligent machines for knowledge production (Kurzweil, 1990) and the human-machine interaction patterns which we observe, initiate and might want to shape also. But, they also concern the social and economic layer of science and knowledge dynamics, the interplay between public and private actors with different (economic) needs and requirements.

So, in the conclusion of this paper we take a turn towards more general questions such as:

- What is the position of archives and other institutions of the GLAM sector which are stakeholders for public knowledge commons in the current race for new AI technologies?
- How can institutions in the public sector be part of the early adopter cohort and this way also co-design technology?
- To which extent do new AI driven Information Retrieval approaches as described in our use case empower both content producers and content consumers?

To address these questions in the last part of the paper, we turn to the model of the Global Open Research Commons (Treloar and Woodfoord, 2024) as a means to reflect on how institutions in the public sector can stay true to principles of Open Science and Research Commons while embracing the distributed nature of the knowledge production process which includes all societal sectors: government, industry, academia and the wider public.

As said above at the core of the paper, we look into engineering explorations to enhance the accessibility of digital collections presented on the web. In those explorations Retrieval-Augmented Generation (RAG) plays a central role. That's why we end this introduction with summarizing the RAG approach, how it relates to LLMs and how it influences the way we think about Information Retrieval.

## 1.1 RAG as a new method in Information retrieval

Large Language Models (LLMs) (Brown et al., 2020) encode knowledge implicitly within their parameters during the pre-training phase, typically by optimizing for next-token prediction. This implicit encoding presents challenges for explicit knowledge inspection, modification, or deletion, thereby limiting the model's adaptability to updated or corrected information (Nie et al., 2025; Guirguis et al., 2024; Wang et al., 2025a; Wang et al., 2025b). Consequently, LLMs are prone to generating outdated, inaccurate, or unverifiable outputs. In certain cases, they may fabricate content—referred to as hallucinations — in order to preserve internal coherence or plausibility.

To mitigate these limitations, the Retrieval-Augmented Generation (RAG) framework has been introduced as an effective approach for addressing knowledge-intensive tasks (Lewis et al., 2020). RAG decouples the knowledge component from the generative model by incorporating two primary modules: a retriever and a generator (Lewis et al., 2020). Upon receiving a query, the retriever searches an external knowledge base for semantically relevant documents, typically through embedding-based similarity measures such as cosine similarity. These retrieved documents, together with the query, are then used to construct an augmented prompt that is passed to the generator—often a pre-trained LLM—which synthesizes a natural language response.

This paradigm offers multiple advantages that address core limitations of pre-trained language models (Gao et al., 2023). First, the knowledge base is decoupled from the model parameters, allowing it to be updated independently, thereby enabling timely and flexible incorporation of new information without the need for costly model retraining. Second, by grounding the generation process in retrieved documents, RAG significantly reduces hallucinations, as the model can rely on externally verified sources rather than internal approximations. Third, the system becomes inherently more interpretable and transparent, since each generated answer can be traced back to specific supporting documents in the retrieval set. Finally, RAG is considerably more cost-effective than retraining large-scale LLMs, which not only require substantial computational resources but also risk performance degradation due to catastrophic forgetting (Wei et al., 2022a). To summarize, RAG builds on existing semantic reference bases, similar to how controlled vocabularies or classifications are more specifically defined than natural language expressions.

Having said this, RAG is not per se easy to be interpreted or transparent to the user. It just provides other ways to check a claim or assertion. Compared to traditional information retrieval systems, RAG demonstrates a superior flexibility by enabling users to issue natural language, contextual, or follow-up questions without predefined templates. The semantic matching capabilities of modern retrievers, combined with the generative capacity of LLMs, facilitate a more robust and user-friendly interaction paradigm for information access and reasoning over unstructured data sources (Tykhonov et al., 2025).

In conclusion, the RAG paradigm provides a more robust framework for building controllable, reliable, and eventually more trustworthy AI systems that can adapt to evolving knowledge with greater interpretability and efficiency. But, different epistemologies continue to exist as expressions of representations of knowledge. To judge the results of the query will remain context dependent and related to the information needs of the user. This cannot be solved by machines but requires the human in the loop, for which those machines are supposed to act as tools for this human in her/his epistemic journey not as avatars taking the place of the human.

While RAG has shown significant promise in addressing knowledge-intensive tasks (Arslan, Munawar and Cruz, 2024; Liu, 2024), it remains an evolving paradigm and is subject to several limitations. One major challenge lies in retrieval quality: the retrieved documents may be irrelevant, redundant, or unhelpful to the current query, and in some cases, a retrieval may not even be necessary (Jiang et al., 2023; Asai et al., 2024; Yan et al., 2024). Additionally, the language model itself may be constrained by limited summarization capabilities or may default to relying on its own internal knowledge (Wu et al., 2025). Furthermore, RAG systems often underperform on complex queries, such as multi-hop questions or those requiring a holistic understanding of the entire document set (Wang et al., 2024; Edge et al., 2024).

To address the retrieval quality issue, CorrectiveRAG (Yan et al., 2024) introduces an auxiliary, lightweight language model to estimate the relevance between each retrieved document and the query, assigning a scalar relevance score. Documents with lower relevance scores are subsequently filtered out, thereby improving the overall quality and utility of the input to the generative model.

Additionally, some other research explored modifications to the RAG workflow, aiming to enhance retrieval quality. ActiveRAG (Jiang et al., 2023) proposes a dynamic and iterative retrieval mechanism, where the LLM actively determines when and what to retrieve based on its internal uncertainty. Concretely, the LLM begins with an initial retrieval and proceeds to generate a provisional sentence. If it detects low confidence in predicting the subsequent token, it triggers an additional retrieval operation specific to that uncertain segment. This sentence-by-sentence decision process continues until a complete answer is formed, thus avoiding redundant or irrelevant retrieval steps.

As a further advancement, SelfRAG (Asai et al., 2024) introduces a more introspective control loop to optimize retrieval decision-making. Instead of retrieving documents by default, the LLM first attempts to generate a response without external knowledge, and then reflects on whether retrieval is needed. If retrieval is deemed necessary, it is performed, followed by a reassessment of the relevance and utility of the retrieved documents. This "answer–think–retrieve–think–answer" workflow ensures that retrieval occurs only when beneficial, thereby reducing unnecessary queries and filtering out unhelpful content.

While RAG systems rely on large language models (LLMs) to generate responses based on retrieved documents, they remain vulnerable to hallucinations and misleading evidence, particularly when irrelevant or low-quality documents are included in the context. To address this issue, RankCOT (Wu et al., 2025) introduces a novel approach that integrates Chain-of-Thought (CoT) (Wei et al., 2022b) reasoning with document reranking and summarization to improve generation quality. CoT is a prompting strategy that encourages LLMs to generate intermediate reasoning steps, thereby enhancing transparency and answer reliability. In RankCOT, this strategy is further enhanced by fine-tuning the model using Direct Preference Optimization (DPO) (Rafailov et al., 2023), allowing it to better assess and prioritize documents based on their relevance to the query. This method effectively filters out irrelevant content and reduces the likelihood of hallucinated or inconsistent responses during the generation process.

Although the vanilla RAG framework performs well on many knowledge-intensive tasks, it faces limitations when addressing complex queries requiring global understanding, such as multi-hop reasoning, multi-turn interactions, or questions that require aggregating knowledge dispersed across multiple documents (see also Meding and Daugs, 2026b). These challenges arise in part due to the limited context window of LLMs and the constraints of implicit working memory, which is driven by the self-attention mechanism. As the number of retrieved documents increases, the model may become overwhelmed, leading to degraded performance and fragmented reasoning (Hołyst et al., 2024; Shi et al., 2023; Upadhayay, Behzadan and Karbasi, 2024).

To address this issue, DeepNote (Wang et al., 2024) introduces an external memory module in the form of a structured "notebook". The LLM initializes this notebook with intermediate facts based on the query. It then performs iterative retrievals to refine and expand these notes by incorporating information from relevant documents. After each round of retrieval, the LLM reevaluates the query, updates the notebook, and rethinks a new query if the facts in the notebook remain ambiguous, until the accumulated facts are sufficient to construct a coherent and accurate answer. This process effectively offloads intermediate reasoning steps to a persistent external memory, mitigating the limitations of the context window.

Similarly, GraphRAG (Edge et al., 2024) targets tasks that require understanding inter-document relationships, such as identifying connections between entities across documents. This method extracts entities and relations from the corpus to construct a knowledge graph, which is then used to identify entity communities through clustering. Instead of retrieving raw documents, GraphRAG retrieves and summarizes these communities to serve as input to the LLM. This graph-based abstraction supports more structured, comprehensive reasoning, particularly in scenarios where an answer depends on synthesizing information from a large and semantically interconnected corpus.

In conclusion, the Retrieval-Augmented Generation (RAG) framework represents a promising direction for enhancing the reliability, transparency, and adaptability of AI systems in knowledge-intensive applications. A growing research has focused on extending and refining the RAG methodology to support complex reasoning and mitigate hallucinations. As a result, RAG is rapidly evolving into a more versatile and robust paradigm, increasingly capable of tackling challenging tasks and delivering answers with higher factual accuracy and contextual relevance.

Parallel with the further scientific development of RAG based methods, its description and evaluation, elements of LLMs together with machine readable KOS systems, continuously find their way into engineering explorations. Whatever becomes available in forms of libraries, but also general ideas on what workflows could look like gets roped in daily innovative practices around exciting services or tools. Hereby, as we will see later, often the development of new standards play a role. Those engineering explorations are often agnostic towards a concrete service, but very much informed by the needs of concrete services.

In our paper we describe a workflow which falls methodologically under the RAG approach, but has been developed in a practice context and more specifically related to a specific archival platform *Dataverse*. Our specific take concerning the discourse summarised above is to describe methods from the perspective of them being implemented, applied, used in concrete research infrastructure settings. We return to more general aspects of new AI based technologies (including RAG) in the conclusions.

## 2. An engineering journey – adoption and exploration of new AI technology (including RAG) in areas of cultural heritage

### 2.1 The case of the MuseIT project and its specific needs

The MuseIT project enabled a substantial portion of the work presented in this paper. MuseIT stands for "Multisensory, User-centred, Shared cultural Experiences through Interactive Technologies" and is funded by the Horizon Europe programme under the call 'Culture, creativity and inclusive society'. Its mission as stated on the website is: 'Aiming for greater equality, democratization, and social inclusion, MuseIT offers technologies designed to enhance and expand access to cultural assets for individuals with disabilities while also contributing to the preservation and protection of cultural heritage in an inclusive manner.'[5] The project type is a *Research and Innovation Action* (RIA). This is relevant in order to understand the composition of the consortium and the kind of knowledge production executed in the project. RIA's are expected to establish new knowledge and/or to explore new or improved technology and/or services and most broadly any solutions. The project is led by the i-School of the University Borås, Sweden, but the consortium brings together research groups from the universities and large-scale research consortiums with Small and Medium Enterprises and organisations in the cultural sector (Ntonga, Pokorny and Olson, 2023). One of the higher education institutions in the consortium is DANS, Data Archiving and Network Services, an institute of the Royal Netherlands Academy of Arts and Sciences (KNAW), where some of the authors of this paper work. DANS' role for the consortium was primarily to contribute its own expertise in fields of research data management and long-term archiving. DANS itself, while part of the institute portfolio of the KNAW, is not a research institute but a service provider for all questions of digital research data. It also hosts a variety of data archiving services for researchers and research institutions (Morselli, Touber, and Scharnhorst, 2025). Being part of the research infrastructure area, DANS participates in research projects such as MuseIT with the aim to stay in touch with new technological developments, being able to do explorations as RIA types of projects offer, all with the ultimate aim to innovate its own services.

MuseIT combined fundamental research with the quest for infrastructural solutions. To enhance the accessibility of cultural assets, finding new ways to represent those assets was very important. Think here of involving various senses in the interaction with cultural assets. Examples are the development of haptic devices, the description of images by text and sound, or the translation between music and other forms of creativity — all examples for multi-sensory representations. Addressing multimodality was an important topic for the project. MuseIT addressed this topic by exploring new technological ways to support co-creation of cultural assets. This way, MuseIT combined both inclusion and better (web) accessibility concerning the consumption as well as in the production of cultural assets. One of the cornerstones of the project was the further development of a platform (JackTrip™) which enabled artists to make music together on the internet time-lag free (Cáceres and Chafe, 2010). Last but not least, MuseIT also aspired to advance the

5 https://www.muse-it.eu/

ways multimodality in the context of accessibility and inclusion could be made part of documentation and long-term archiving. More concretely, the project plan formulated, "One of the key outcomes of MuseIT is the content repository capable of indexing, storing and retrieving multisensory, multi-layered representation of cultural assets…".[6] This is where the experiences at DANS both with consulting on standard archival techniques as well as with pushing the boundaries of such archival services became relevant.

For both research as well as documentation in the MuseIT project[7], the ways to order knowledge played an important role. A so-called Knowledge Organisation System (KOS)[8] encompasses taxonomies, ontologies, classifications, thesauri, etc.. For collections, the metadata schemes used in indexing the collections are one example for a KOS. But, in each scientific field KOS are also used to annotate the scientific content, representing the canon of relevant and connected concepts that define the heart of a field. In the discourse around Knowledge Organisation Systems, it is not always made explicit that KOS have different epistemic functions. In MuseIT, a lot of ontology engineering took place to develop *new* ways to conceptualise multimodality. While existing KOS were reviewed and taken into account during this scientific journey, in this particular part of the project, the emphasis was on the development of new ways of thinking and categorizing. KOS in such a heuristic function are by default not primarily interoperable (Scharnhorst and Smiraglia, 2021). If one thinks about the use of KOS for indexing content, interoperability is part of making content or data FAIR (Wilkinson et al., 2016). FAIRness of data (or more broadly, any kind of content) is a precondition to scale up information processing operations across various collections of data. Interoperable here means machine-readable in a way that semantic meaning is also transported. But, even in this area of documentation if it comes to the use of KOS, there is always a tension between KOS tailored specifically to the needs of a collection, their curators and their designated user communities, and the request to being 'linkable' across various bodies of knowledge as represented in various collections. This is not a trivial task (Gueret et al., 2013) as much of the work on KOS registry, catalogues and cross-walk workflows can testify (see e.g., Nyberg Åkerström et al., 2024). Inside of MuseIT we find both: ontology engineering as a means to formalise and extend our way to think about inclusion, accessibility and multisensory representations (see e.g., Domingues et al. 2024). But, we also find implementation of KOS in indexing collections, combining existing and *new* aspects in KOS when it comes to multimodality (Tykhonov and Olsen, 2023).

In MuseIT, one of the most recent experiments concerned an implementation of a RAG method in the form of a "Chatting with collections" local chatbot. In this workflow, as typical for RAG (see above) LLMs (sometimes called unstructured data) are combined with KOS (sometimes called structured data). But before one can construct such a chatbot a couple of steps were needed: First to have a repository or archive, second to apply

6 Project proposal, personal communication Scharnhorst.

7 And yes these are two different practices with different norms, values and different skill sets needed in the execution – see (Daga et al. 2023) for more details.

8 KOS is a technical term well defined in the field of Knowledge Organisation, which is situated at the cross-roads of philosophy of science, information and computer science. For the International Society of Knowledge Organisation see https://www.isko.org/ko.html (Accessed November 6, 2025).

suitable KOS for its content, and only then as a third step an experiment to locally implement a chatbot could happen. The next section describes those steps around the MuseIT repository approach (Kontogiannis et al., 2025). We start with the choice of an archival platform to then turn to a specific instance of such an archive on which various experimentations were executed. Describing this engineering journey, we also highlight aspects of the nature of this practice, the preconditions created by project financing, the skills needed and the discussions which have taken place.

## 2.2 An engineering journey – from 'everyone can archive' to a locally implemented RAG

### 2.2.1 Everyone can archive

MuseIT very boldly stated in its project workplan that there is a need for a cultural heritage archive for storing multisensory representation. Talking about a repository or archive immediately includes questions of what collection(s) of digital assets should be archived, how to order them and make them accessible for users, and how to adhere to requirements for digital (long-term) preservation. Institutions rely on various approaches. Sometimes they have the capacity to build tailored software solutions in-house, sometimes they rely on commercial services. But, most often ICT capacities to build tailored solutions are not there. This holds for institutions in the higher education sector as well as for institutions in the GLAM sector. A middle way is to rope in the expertise in archiving of institutions operating in the public sector, and to use software which is developed and maintained by open software communities. This is the approach, partners in MuseIT have chosen, and the project provided the means to support consortium members in taking the first hurdle in setting up a digital archive. DANS, as a member of the MuseIT consortium provided this support and eventually the choice for a digital archiving platform went to *The Dataverse Project* (Crosas, 2017). Dataverse is an open source web application to share, publish and explore research data. Developed first at the Institute for Quantitative Social Science at Harvard University, there are currently about 137 installations of Dataverse worldwide with many collections (Dataverses) and datasets. DANS has built its own archival data services for the Dutch higher education institutions (DataverseNL[9]) and for Dutch researchers directly (DANS Data Stations) implementing this software. DANS engineers have also actively contributed in the further development of the software, and – drawing on the funds provided by other European projects – developed ways to make setting up a Dataverse instance easier. This has been described as 'Archive in a Box' (as described in Wittenberg et al., 2022). This significantly lowers the threshold in setting up a repository – so *Anyone can archive*. This is very similar to the ways one could observe how information visualisation has been democratized over the last decade. For the latter see for example the *Places and Spaces*[10] long-term project of Katy Börner as documented in the series of atlases of information

9 More information available at dataverse.nl (Accessed November 8, 2025).

10 More information about the *Places and Spaces: Mapping Science* enterprise can be found at the website: https://scimaps.org/home (Accessed November 10, 2025).

visualisation (Börner, 2010, 2015, 2021; Börner, Record, Theriault 2025) of which the second carries the title *Anyone can map* (Börner, 2025).

The tasks MuseIT wanted to be supported by one or multiple repositories were multifold. Among them was:

1. to have a shared repository during the project lifetime to be used across all work packages;
2. to choose a technical solution which is flexible towards the needs of various collections which emerged in the various use cases of the project (examples are cultural heritage material, but also scientific articles, and ways to directly save data streams from sensors);
3. to have a place to register various ontologies and KOS relevant for aspects of accessibility and multimodal representations;
4. to have a repository with metadata scheme(s) which can be tailored towards the variety of content mentioned above and
5. a variety of means to interact with this content.

Needless to say that the repository should operate on the web, and adhere to FAIR principles concerning machine actions. At the end MuseIT implemented two instances of Dataverse: a general MuseIT Dataverse instance, to manage all project related research datasets. To this general instance partners contributed with datasets, but also by designing the metadata schemes and developing workflows around the instance (Kontogiannis et al., 2025). A second more specifically tailored instance was developed by ShareMusic[11] – one of the cultural heritage partners in MuseIT together with DANS. Around this instance several experimentations involving AI based methods were executed – as we will sketch in the next two subsections.

To summarise, there were several advantages to choosing Dataverse as a software solution. The software project itself operates in the public sphere and adheres to Open Science principles. The repository solution, as we will see, is flexible enough to be both interoperable concerning core standard information on datasets, as well as to allow designing a collection specific, flexible metadata scheme. The existence of a Dataverse global engineering community secures the release of an updated, tested *master version* while at the same time creating a forum for experimenting with the newest technological possibilities.

Concerning AI methods, the Dataverse community as Boyd presented recently currently has more than 10 workflows developed (Fig. 1), which include AI components (Boyd, 2025).[12] Some of them are production ready, some are still in development. In some of them (e.g., *Croissant* (Akthar et al. 2024), and the Model Context Protocol) DANS

11 ShareMusic & Performing Arts is a Swedish national knowledge centre for artistic development and inclusion. For more information see their website: https://www.sharemusic.se/en/home (Accessed November 10, 2025).

12 https://dataverse.org/presentations/ai-and-dataverse-project-using-ai-benefit-research-data-community-members

has been involved. As we will see, some of those ideas have also found their way into the MuseIT project.

*Figure 1: Overview of AI related workflows in Dataverse – Slide reproduced from Celyn Boyd's presentation September 2025 (Boyd, 2025).*

Dataverse AI-Related Resources, Tools & Projects

| Resource, Tool or Project | Type | Improves | Status |
|---|---|---|---|
| AI Guide | Resource | User experience | Production |
| Ask the Data | Tool | Data reusability | Production |
| Ask Dataverse | Tool | User experience | Experimental |
| AutoSage | Tool | Metadata quality, User experience | Experimental |
| Croissant | Tool | Data findability | Production |
| **Enhancing Dataset Metadata Project** | Project (Tool) | Metadata quality, User experience | Development |
| GREI AI Taxonomy | Resource | Understanding of AI roles in RDM | Production |
| **Research Data Metadata Knowledge Graph** | Project (Tool) | Data findability, Data reusability | Development |
| Model Context Protocol (MCP) Server | Tool | Data findability, Data reusability | Production |
| **Spam Detection & Workflow Automation** | Project (Tool) | Repository data quality | Investigation |
| TurboCurator | Tool | Metadata quality, User experience | Production |

Dataverse dataverse.org @IQSS Ceilyn Boyd

## 2.2.2 Customize your archive – build your own knowledge base

In the previous section we already referred to the objectives of MuseIT. These include to increase the accessibility of cultural assets by the development of multisensory representations, enabling engagement by the public regardless of functional or sensory impairments, as well as to develop methodologies for preservation and safeguarding cultural heritage assets. For the creation of representations and the final design of repositories Knowledge Organisation Systems (KOS) are key. KOS are used in defining the metadata describing the cultural assets, but also the ways in which one can interact with those assets.

One of the consortium members of MuseIT, the organisation ShareMusic, came with a specific need concerning a repository. This Swedish cultural organisation describes itself as "a national knowledge centre for artistic development and inclusion" and states on its website: "We work for everyone's right to experience, participate in and practise artistic and cultural activities".[13] ShareMusic organises events of performing arts next to workshops and seminars. All those activities are dedicated to shaping a "new cultural landscape with new expressions and narratives by disabled practitioners"; to "increase[d] knowledge of inclusive practices across the cultural ecosystem" and ultimately to contributing to a richer and more equal cultural life in Sweden — and the world — that reflects the diversity and richness of society".[14] These are just some expressions of the aspirations of this organisation. What makes ShareMusic also special is that to realise these goals the organisation has a long-standing tradition to work together with scientific and other knowledge producing institutions to test new methods to enable both production and consumption of cultural assets, here music, despite any sensoric impediments of

13 Quotation from the website: https://www.sharemusic.se/en/home

14 See Footnote 13.

involved actors. In consequence, their 'collection' of material is wide and varies from scientific articles (as in a classical digital library), to performance recordings and other live data streams which might deserve a long-term preservation, and to newsletters and blog posts about performances. The latter represent ephemera which are often not preserved because of their temporary character. Those traces – if by chance preserved – enable a richer look into the past. To summarize, ShareMusic is an organisation with one foot in the cultural sector and one foot in the innovative processes as executed in academia and other *Mode2* knowledge organisations with a collection of material at hand which is varied by nature.

The aspirations of ShareMusic when designing *an archive* for this material went far beyond having just a place to index and preserve this material. As detailed in (Johansson et al., 2025), ShareMusic wanted to build a Knowledge Base for Art and Inclusion, which conceptualisation embraced new KOS to describe features relevant for accessibility and inclusion. This knowledge base should inform the curation of the material and make the work of those curators easier, but should also enable various different interfaces to interact with the collection on-line.

Again the choice to rely on Dataverse as a technological solution to archive and index these digital traces (data) proved right. Part of the agility of this specific infrastructural approach to archiving lies in the possibility to extend the metadata scheme which usually comes as a standard block for any Dataverse instance. Not only can the metadata schema be extended, it is also possible to automatically connect it to other controlled vocabularies relevant for the specific content of the ShareMusic collection and the specific interactions with this content which have been envisioned. Such connections had been tested for Dataverse in other projects, so workflows were available and could be adapted to this specific case (Myers and Tykhonov, 2023).

At the end, for the ShareMusic instance, Dataverse acted as an 'invisible archive', as a technological backbone complemented by interfaces to support curation and access. As visible in the snapshots of the website[15] below (see Fig. 2) the modalities in which material was digitally documented and ways of interaction were also indicated. During the explorations, further AI supported workflows, e.g. to describe a picture with words, or to sonify an image were tested. The final implemented solutions are rather standard, and the overall accessibility of the web-based interface is not yet fully developed. But, the specific requirements of the organisation and the current implementation of an archive with a set of new features proved an interesting learning experience for both engineers and those working in such a cultural institution.

As Fig. 2 shows for the ShareMusic Dataverse instance an additional metadata block has been designed containing fields as modalities, ways of interaction, and art form. This block also contains further ShareMusic specific elements of knowledge ordering as *Topic*. Using the *Topic* feature, and its subcategories *Topic Name* and *Topic Type*, ShareMusic's data curators are able to create sub-collections around events and/or projects. Those orderings are visible on the *Knowledge Base* interface, which is the envisioned main entrance to engage with the content (see Fig.3). In this specific interface, other metadata features

15 For the Dataverse instance used as backbone see: https://database.sharemusic.se/dataverse/root?q=&types=dataverses%3Adatasets&sort=dateSort&order=desc&page=1

as for instance modalities are emphasised as ways of interaction. It is also this interface, in which a local AI chat bot is hidden behind the toggle *Ask Our Assistant*, the features of which we will describe in the next section.

*Fig. 2: Snapshot of the Dataverse instance at ShareMusic, highlighting the added metadata fields.*

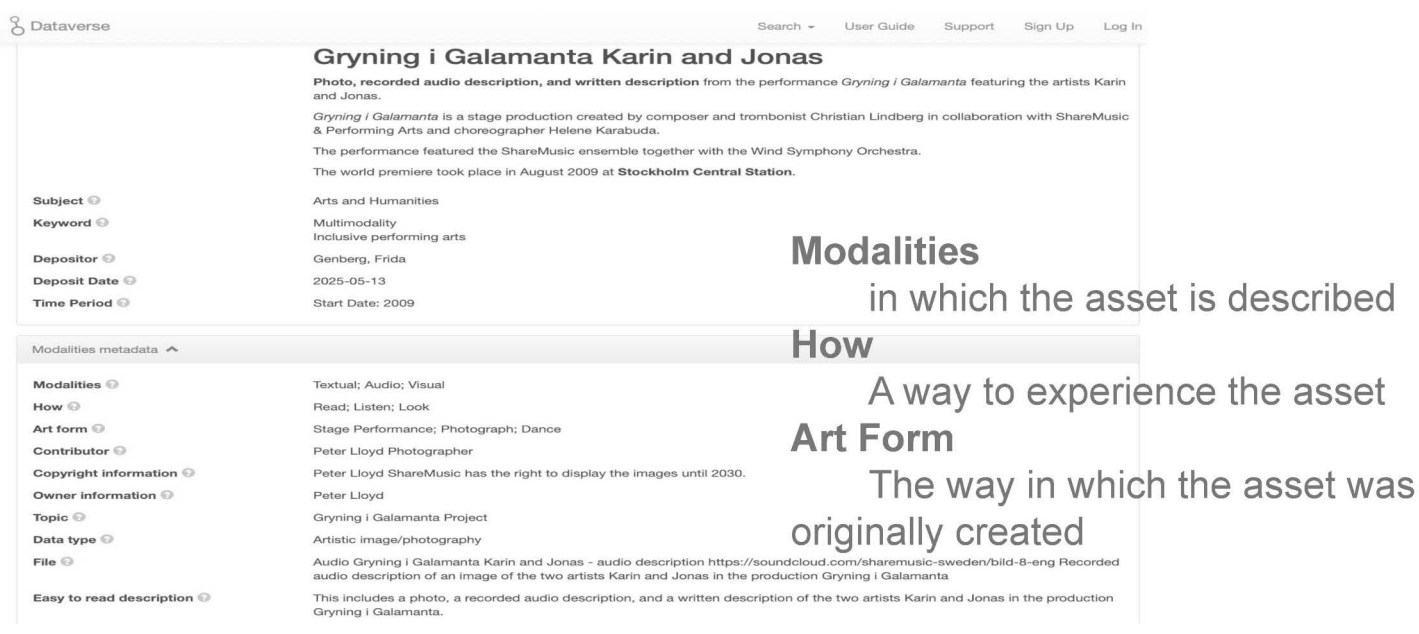

*Fig. 3: The Knowledge Base designed for the ShareMusic collection*[16]

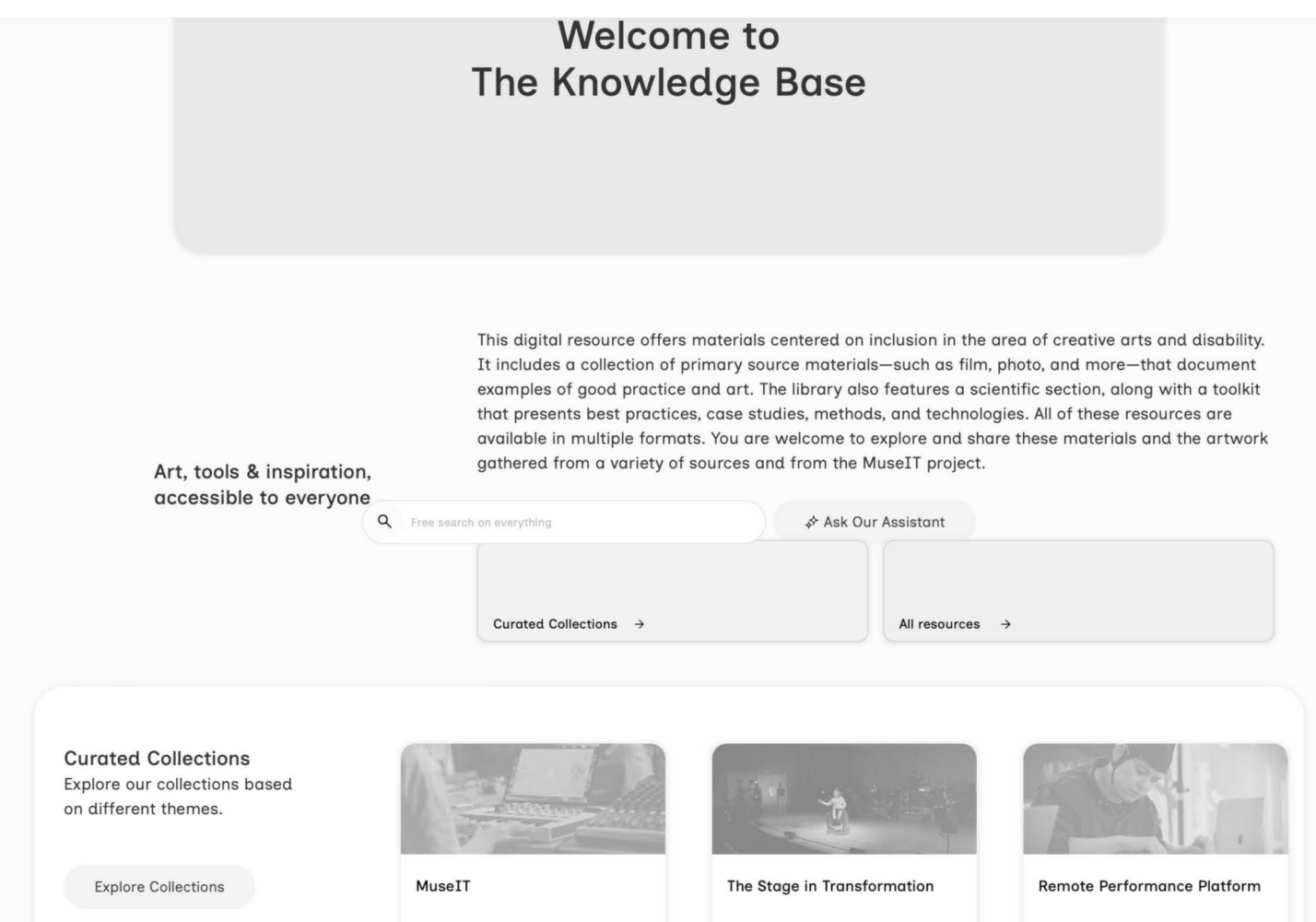

16 Snapshot of the website https://knowledgebase.sharemusic.se/ – the interface to interact with collections (Accessed October 18, 2025).

## 2.3 Chatting with a collection – LLMs and RAG tested in practice

As described in the previous sections, the engineering journey which led to testing LLMs and more specifically RAG approaches for a concrete archive started with (1) enabling to install archival (open source) software for a wider group of institutions and continued with (2) various explorations to integrate new technological solutions into this archival platform solution. An inbetween step was to enhance the metadata description of the archived content. Here, possibilities to integrate links to other existing machine readable KOS as part of the indexing was one important step (De Vries et al., 2022, Zamborlini et al. 2024). The currently existing universe of linked KOS, which is part of the semantic web technology (see e.g., Merono Penuela, 2016), fosters reuse of KOS and standardisation, but it also adds meaning to content on an unprecedented scale. This – as shown in the short literature review above (see the section Introduction) – is the precondition for any RAG approach. For an institution as DANS which provides archival services, the task to make collections better accessible, and to foster re-use is key. One way to enhance the potential of re-use is to convey meaning to the deposited datasets. How to do this is a very intriguing problem. Semantic meaning is traditionally codified in the metadata scheme models and the values of their parameters. But, it does not translate automatically and easily to a wide range of users. As we pointed out in the introduction, KOS are always specific – meaning here that they represent a distinct specific framework to represent material which evokes specific types of reflections. For the DANS Datastations, which are for good reasons discipline oriented, adding discipline specific information to the metadata is a long standing issue.

When the new generation of Generative AI emerged, the question immediately arose which of those features could be used also for a better Information Retrieval around Digital Archives. In our own internal discussions which were also interdisciplinary by nature, we were looking for metaphors to differentiate between a KOS and a LLM approach. In the first approach meaning to the content would be given through connecting towards various KOS (also sometimes called structured data). In the second approach meaning would be extracted from the content itself directly using emerging patterns and in the case of LLMs this would be textual ones. One metaphor we used in our internal communication was to see LLMs as experts which one can ask in natural language. This would be different from relying on KOS which we saw similar to asking a librarian, who while not familiar in depth with all content, is apt in the meaningful ordering systems for types of content. At the end, the workflow for the implementation of a local chatbot together with the web interface was named *Ghostwriter*. Fig. 4 (as presented by Mayr et al. 2025) gives an overview how the Ghostwriter approach relates to other Information Retrieval approaches.

*Fig 4: Position of the Ghostwriter approach in relation to other information retrieval methods. Slide from the presentation (Mayr et al., 2025)*

The *Ghostwriter* approach - part of the Retrieval Augmented Generation in Information Retrieval

First challenge to solve a problem: find the right question?
(and right person or information source and interpret the results)

| Query | One database representation<br>Me and a database | Need to know the schema and its usual values to get a result | Classic Information retrieval problem: find the right query |
|---|---|---|---|
| Query | One data(collection/space)<br>Connected structured data(bases/graphs) in the background<br>Me and a librarian | The machinery gives back suggestions for possible similar/better queries based on the connection between schema's<br>And a list of possible results for each of the variants | Google Feature and schema.org; machine makes informed guesses about your query<br>Works on the web, not on a more local interaction |
| Query | Large Language Model<br>Me and a library; Me and a round of experts | Interprets the query as a natural language input, and suggests results again expressed in natural language | |
| Query | LLM (local) + target data(collection/space) + embedded in a network of additional data interpretation sources (via API's)<br>Me chatting with experts and librarians at the same time | Creates a family of terms around the query; identifies related structured information;<br>Returns a list of results | If applied iteratively can help you to reformulate your question by getting a better understanding what do you actually want to ask and what can you actually ask towards the available data space |

Slides from: https://doi.org/10.5281/zenodo.17242902

The precondition for the *Ghostwriter* workflow to function is to express all available information (from the collection in question, but also all contextual enrichment one can connect the collection to on the web) in a machine-readable form ready to be processed. This other workflow preceding *Ghostwriter* was called internally *EverythingData* (Fig. 5).

*Fig 5: Schematic overview about the relationship between EverythingData, Ghostwriter and RAG in general (Slide from Mayr et al., 2025).*

Ghostwriter and EverythingData
Wider Discourse - Retrieval Augmented Generation (RAG)

**Main Ingredients**

**Vector space** constructed from the *content* of data files encoded in embeddings, with properties and their attributes. Embeddings are computed by various ML algorithms and use different LLM's.

**Graph** represents a *metadata* layer integrated with various ontologies and controlled vocabularies including responsible AI. Graph is expressed by Croissant ML standard.

**Vision**: Let's get both graph and vectors together in one model (GraphRAG). Let's do this 'locally' as a kind of *Distributed AI*, where the LLM is "interface" between human and AI, and works as a "reasoning engine".
**Implementation**: LLM connected to the "*RAG library*" (graph), navigate through datasets and consuming embeddings (vectors) as a context.

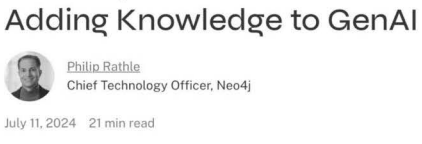

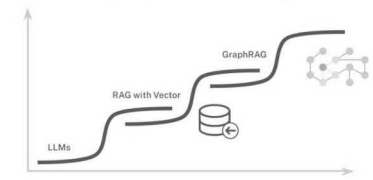

https://neo4j.com/blog/genai/graphrag-manifesto/
https://en.wikipedia.org/wiki/Retrieval-augmented_generation
(see Arno Simons Presentation: tool box)

*Fig 6: Schematic overview about the Ghostwriter workflow (Slide from Mayr et al., 2025).*

Ghostwriter
and EverythingData

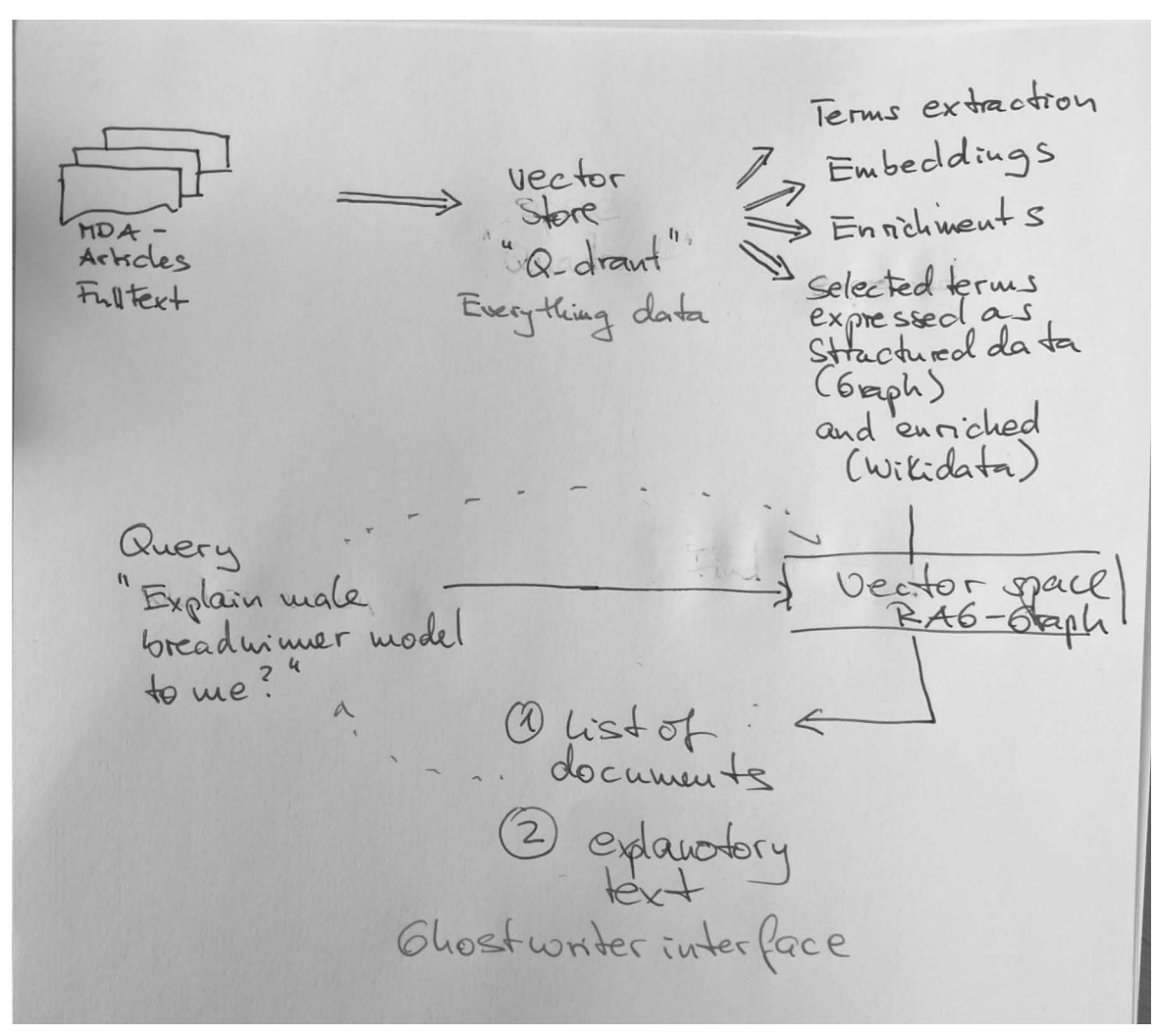

The *Ghostwriter* workflow entails an interface that enables the user to ask questions in natural language. Fig. 6 gives an schematic overview of the workflow. In this figure (Fig. 6) the workflow has been applied to another application case: namely a collection of articles from the *Methods, Data and Analysis* journal (for more details see Tykhonov at el., 2025). The query result is presented as a natural language summary together with a list of relevant hits. The innovation for both LLMs and RAG is that the presented list of results does not rely on term or phrase matches only, but that the suggested sources resonate in some wider sense – either based on patterns in natural language (LLMs) and/or based on concepts expressions (RAG) – with the query. Based on complex stochastic models the idea is to better capture the semantic meaning of the original question, and suggest related sources. Ghostwriter has been tested for various collections. Among them are the webcollection of the Now.Museum (Vion-Dury et al., 2023) and (as mentioned above) a test collection from a paper archive of the journal *Methods, Data and Analysis* (Tykhonov et al., 2025).

The design of the MuseIT ShareMusic repository offered another possibility to test *Ghostwriter*. In this case, as visible in Fig. 3, the Knowledge Base Interface contains a button "Ask our Assistant". This button can be found next to the usual area where a search query can be entered using free text. When clicking on this button, the user is guided to another interface (see Fig. 7). Here, another free text field is available for asking a question about the collection. Fig. 7 displays a snapshot for the question "which performances does the collection contain?". This search once executed leads to a refreshed website which contains an answer given in natural language (Answer section), and a section (Sources) in which a number of most relevant datasets from the collection are listed. In its current form, the interface allows the user to re-iterate their question, while the response always comes in a combination of natural language and lists of hits. This way, the user is invited to explore a collection, rather than being guided in a seamless way.

*Fig. 7: Example for the use of the local chatbot on a collection – ShareMusic case*

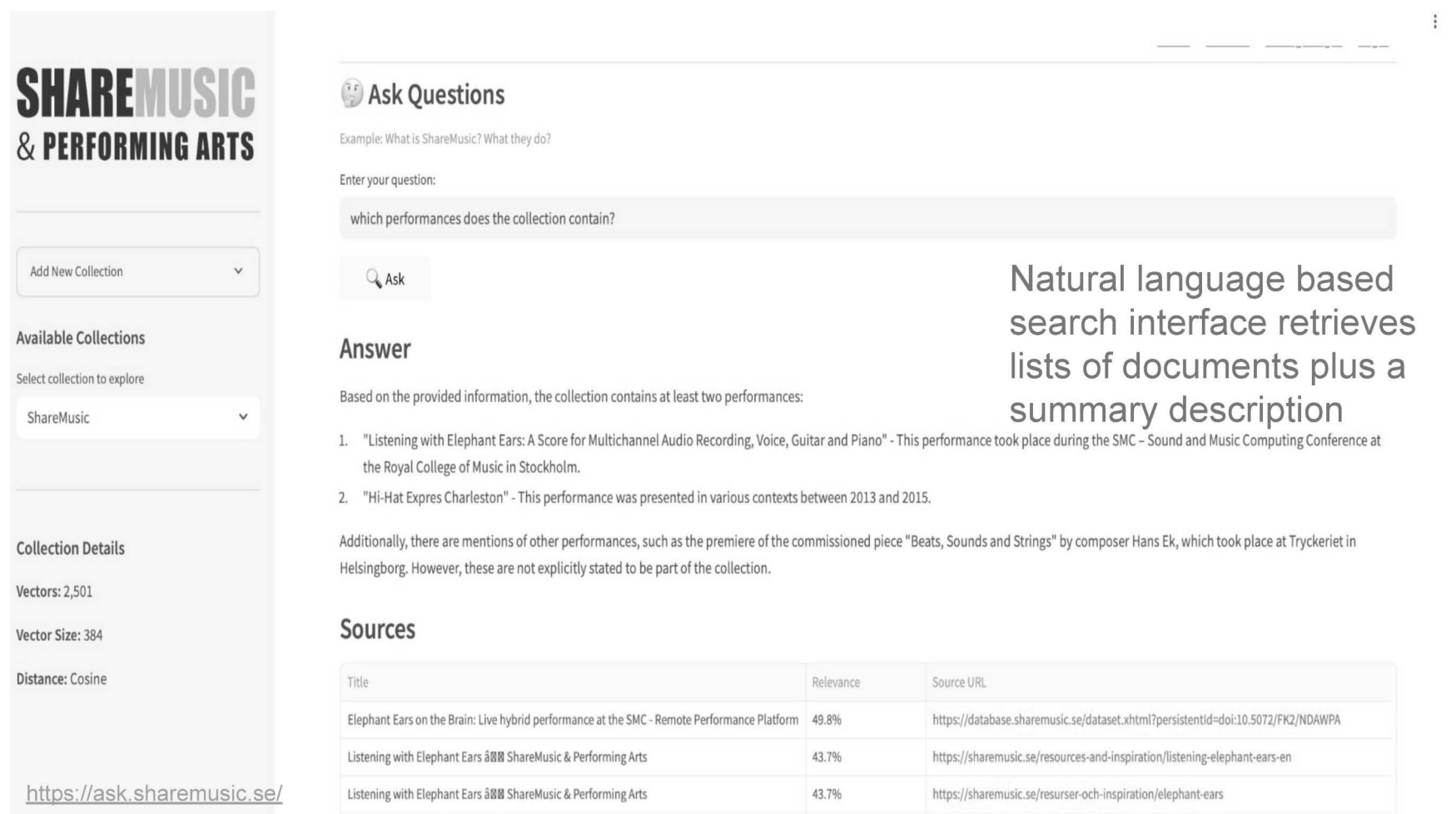

This form of an AI agent can meanwhile be found in many web-based information service systems, the most known mundane one is possibly the AI assistant in Google Search. As we will touch upon in the conclusion, this cannot come as a surprise, if one looks into the collaboration pattern among engineers providing services (commercial and public) for information navigation. The interesting aspect of such 'chatting with …' approaches for a collection in the cultural (heritage) sector, is that the chatbot runs locally. Although the *Ghostwriter* workflow – as described above in the RAG literature review – relies on ready-made LLMs, there is a possibility to use other LLM models developed under greater transparency and with a possibility that such LLM models are also being trained locally. To which extent such 'AI assistants' will be embraced by users, and meet the interaction needs, will have to be studied scientifically. But, the engineering journey in the MuseIT case shows that public sector institutions do not need to wait until commercial solutions are ready. On the contrary, with their specific and often 'out of the box' requirements they play an important role in co-shaping such new forms of technology. We discuss this a bit more systematically in the next concluding section.

## 3. Conclusion

The exercise of developing the repositories for MuseIT, complete with chatbot for innovative search functionalities, shows on a more abstract level how public knowledge infrastructures can benefit from *trading zones* (Gallison 1997) where a variety of actors (scientists, heritage professionals, engineers, etc.) come together and co-create technology, while reflecting on its meaning and function at the same time. In this conclusion we reflect about the process "Co-creating AI technology'. Here we rely on a specific conceptual model which formalises how actors collaborate in knowledge and research infrastructures, and which allows us to discern the elements necessary to create such a trading zone. This model is the so-called Global Open Research Commons (GORC).

While MuseIT is a specific case, with specific needs, it is also typical for the use/adoption of new technologies among institutions which operate at the boundary between academia and the wider public. As we flagged out before some of the MuseIT partners constitute institutions and services which are part of the research infrastructure. Research infrastructures are often conceptionalised as facilities, instruments, services provided to support research. What is less in the focus are the institutional carriers of those functionalities. Despite the lack of a shared definition of a research infrastructure (Morselli 2023) there is a wider discourse about research infrastructures and a quest for a terminology or model to capture various aspects of its function. One of these models is the Global Open Research Commons (GORC) (Treloar and Woodfoord 2024) developed by an Interest Group of the Research Data Alliance over the last couple years, and meanwhile applied in various contexts (Treloar, Woodford et al. 2025). The motivation to develop the GORC model emerged from the concern to keep scholarly outputs (publications, data, software and workflows) as *commons* – public goods in a situation where knowledge production in and outside academia are genuinely interwoven. As we will use some concepts of this model to reflect on the questions put out in the introduction, let us first give a summary of the model.

### 3.1 Global open research commons – a lens to look at the adoption of technological innovation

The GORC model is based on a typology of nine essential elements which all need to be in scope when considering to what extent a research infrastructure service, operation of an institution, or a network activity is devoted to Open Science and the idea of Research Commons. Consequently, it can be applied both as a framework for discussion as well as a non-prescriptive 'check-list' of possible considerations. As a heuristic device the model itself is in ongoing development, and the position, names and detailed definition of elements and their relationship will change over time. Still, and this makes it so suitable for this conclusion section, GORC reminds us in a very elegant way that knowledge production combines epistemic and social-communicative processes; that research infrastructure is never only about provided services; that technological innovation is about new ideas, but also about social, communicative, economic and political processes if it comes to the adoption of a technological innovation. The more profound the technological change, the more important these mutual dependencies.

The GORC model is best summarised in the visualisation below (Fig. 8). All elements shown in this schema are high level concepts necessary to be addressed when enabling Research Commons. The five elements placed in the upper part of the visualisation (with a white background) represent the norms and values to be followed when constructing a truly research commons. In them, the human factor prevails. The three elements (in blue) below represent the ways in which interaction with the research infrastructure takes place, and their nature is more technically defined. The middle element *Standards* (in dark blue) stands for a needed connector, coordinator between the norms, values, means and the technical feature. All elements together are needed to create a research commons.

*Fig. 8: Essential Elements of the GORC model*

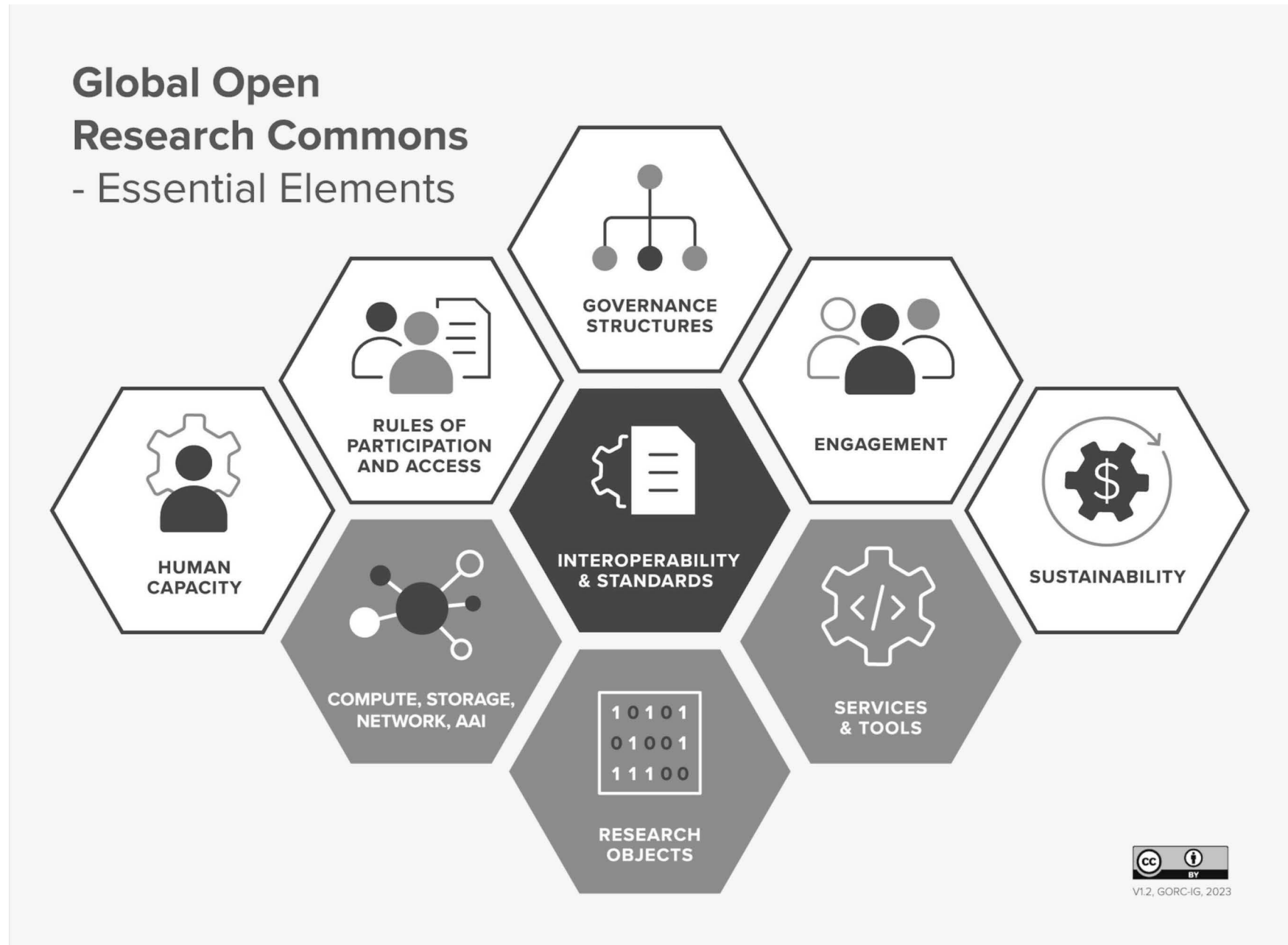

Similar to the concept of *knowledge infrastructures* (Edwards et al., 2013) which a decade earlier shed light on the collective nature of knowledge production and the necessary embeddings for knowledge production to happen, the GORC model highlights the necessity for interwovenness between soft skills or the human factor and the technological aspects down to their tangible manifestation in machines (Treloar et al. 2025).

The model has been already tested by network organisations such as the European Grid Initiative (EGI, https://www.egi.eu/), or national organisations providing research infrastructure services as SURF in the Netherlands, and CSC in Finland. But, it is currently also used to coordinate the establishment of various national and thematic nodes for the European Open Science Cloud (EOSC). Beyond its use in such rather institutional settings, testing the maturity of methodology might be another field of application. New workflows and tools as presented in these proceedings once they become standards and sometimes *services* (in a research infrastructure sense) come with additional requirements which can be described using elements of GORC such as Engagement, Skills and training, or Governance and Sustainability.

But, the GORC model can also be useful to reflect about how projects oblige to Open Science and Research Commons norms. It complements in a way the already required data management plans by adding necessary ordering principles of the processes leading to Open Science. Returning to our concrete use case, the MuseIT project, we find many GORC elements being present in the process. From the networked nature of knowledge production follows that not all elements relevant for research commons are *in one hand*. Different organisations might take care of various of those elements. But only if they interact together based on clearly defined rules – as e.g. represented in standards for interoperability, a research commons can truly emerge and issues of sustainability can be addressed.

In MuseIT, we see scientific institutions and SMEs taking care of specific aspects of the more technical elements. They partly execute fundamental new research for these areas and partly take care of the implementation cases. DANS as a research infrastructure organisation delivered consultancy for a variety of those elements. The work of other organisations – like ShareMusic – could be best located with respect to elements such as Engagement and Sustainability. Part of the success of MuseIT's aspiration concerning setting up a new kind of repository relied on the re-use of technology already developed (Dataverse) which in itself addresses elements of research commons. In particular, the use of standards such as *Croissant* (Akthar et al. 2024) as part of the Dataverse production system, created the foundation on which experiments with RAG could be executed.

## 3.2 Position of archives and other institutions of the GLAM sector in the current race for new AI technologies

As has been stated very clearly in the review paper of Simons et al. (2026) there is a clear divide in terms of skills between those who build LLMs and experiment with AI technologies and those who 'apply' ready-made tools. As the auhors express, "[i]n this landscape, LLM literacy becomes essential: not just technical skill, but interpretive awareness of how models work, what assumptions they encode, and how they shape inquiry. These issues also raise broader questions of accountability and the politics of research infrastructure". A recent report of SURF on Technological Trends (SURF 2025) explicitly names the lack of 'prompt-engineering' skills as an issue that needs to be addressed. This is nothing new. Already concerning general computational skills it remains important that all participants in building research commons have a basic understanding of fundamental principles on which machines operate. Looking into the history of AI debates can be very enlightening here (e.g., Kurzweil 1990). But, there is also no going back to a situation where everybody involved is a universalist. Current bodies of knowledge are very differentiated, detailed and far too large for one person – even with the help of machines – to overlook. Division of labour or skills and collaboration have been answers for such situations, and the application of AI/LLMs is no exception for this. But, we also know from studies of interdisciplinary work that more is needed to make collaboration effective: (1) arenas need to be created in which actors with various skills can meet and (2) a terminology needs to be found to enable communication across epistemic borders. In other words *trading zones* are needed (Gallison, 1997).

If such *trading zones* are not proactively organised, in the current AI debate various actors naturally will address those issues which fall into the realm of their own competence. For research infrastructure support institutions as DANS, the designated communities for which DANS develops services rightly dominate the discourse. But having dimensions of mature services as a leading principle does not mean one can not engage in the shaping of technology itself. Archivists have never been passive executors of community needs (Borgman, Scharnhorst and Golshan, 2019); GLAM institutions become strong in the adoption process of technological innovations when being able to co-create[17] (Rodighiero et al., 2022).

We know from innovation studies (Rogers 1962) that in the phase of technological change one always observes the phases of Early, Mid and Late Adopters. We know from the same field, that once a technology is ripe and has won a dominating place, a lock-in often is formed and alternatives stand no chance (Bruckner et al., 1994). This is why it is not only a *nice to have* but a necessary condition / *sine qua non* for adopters which epistemically might be farther away from the 'fire' where the technology emerges, to mingle, play, experiment with new technologies. Only by experiences in practice, chances of co-creation and co-shaping technology emerge.

Project funding such as the one enabling MuseIT are the socio-economic bases for a *trading zone* around new AI technologies and their application and use in GLAM settings. Looking at the MuseIT case of co-creating AI through the lens of the GORC model reveals which other ingredients beyond project funding are needed. The MuseIT consortium brings various stakeholders together from academia, (small) industry, and the cultural sector. As said above those stakeholders take care of various of the elements needed to produce research commons. They do this bringing their own expertise, infrastructure and past project results to the table. For DANS, the explorations around AI methods used for archives are part of a longer and broader engineering journey (see previous section). This is a journey that started with enabling a wider adoption of archival software standards (in this case Dataverse), while at the same time updating this same software with new methods, technologies – from machine-based linking of KOS specific KOS to the larger KOS universe to the local installation of a trained chatbot (an AI assistent).

Looking at the path-dependencies in the development of Dataverse itself we also see links to movements like agile or pop-up digital data infrastructures (Tykhonov 2020) which were developed during the COVID pandemics. But, there are also links to current incubator networks such as the *ML Commons* (https://mlcommons.org/working-groups/data/croissant/) which brings together engineers from industry and the public sector with computer scientists and engineers located in academia to explore new standards – like the *Croissant* standard. These new alliances acknowledge the emergence of what sometimes in the history of the AI discourse has been called a 'science-engineer' making a metaphorical reference to the artist-engineer of the renaissance[18] (Haug et al., 2012)

The GORC model reminds us how in the process to achieve research commons – actually for any type of knowledge commons – technical aspects, organisation of workflows and embracing of communities come together. In the case of the Dataverse in-stance for ShareMusic in the MuseIT project, the requirements of the organisation to organise, archive, preserve and make accessible of their own collection were leading. The KOS features the organisation eventually chose had to be practical for the working of the guardians and curators of the cultural assets, but also practical for the interaction with the material. Given the specific communities with and for which this organisation works, information retrieval tools had to adhere to new forms of interacting with material. These requirements influenced the search process for which type of AI supported method in multimodal representation would be interesting to be tested. This, as described above, could be the description of an image or the reading of a text. The implementation of a local AI assistant extends the possibilities to 'explore' the content of the collection. As pointed out in the RAG literature review, RAG supports an information retrieval paradigm in which the empowerment of the user is central. Instead of seamlessly paving the way – the reference systems, the KOS as ordering principles, remain visible. The user is guided but not blind-folded, he/she is invited to explore on their own terms.

---

17 See as an example the AI Explorer of the Harvard Museum https://ai.harvardartmuseums.org/about (Accessed November 20, 2025).

18 Quotation from Historisch-Kritisches Wörterbuch des Marxismus, Bd. 8/1, Künstliche Intelligenz: [Dass KI] "einen neuen Typ von *Wissenschaftler-Ingenieur* hervorgebracht hat, vergleichbar dem *Künstler-Ingenieur* in der Renaissance (RAMMERT 1995, 14), gilt nicht nur für die Ingenieurselite. TURINGS revolutionäre Leistung wurzelt im verallgemeinerbaren Aspekt seiner geschichtlich neuartigen Integration von Hand- und Kopfarbeit, van Theorie und experimentierendem Eingreifen in Maschinerie." Historisch-Kritisches Wörterbuch des Marxismus, Bd. 8/1, Künstliche Intelligenz. For the comparison to the Renaissance see also Zilsel (1942), Rossi (1962), and Long (2011).

Let us at the end of these conclusions return to the questions we posed in the introduction and present possible answers based on our experiences.

- What is the position of archives and other institutions of the GLAM sector which are stakeholders for public knowledge commons in the current race for new AI technologies?
    - Archives and other institutions in the GLAM sector are excellent co-creation partners in shaping new AI technologies. They adhere to principles of public knowledge commons and with their specific function come requirements which often pose new puzzles to the engineering process. Practice is eventually the test for any new technology, and the cultural sector often has played a pioneering role due to its variety of needs and practices.
- How can institutions in the public sector be part of the early adopter cohort and this way also co-design technology?
    - To be able to become a partner of equal rights, arenas are needed where co-creation can take place, and trading zones between different epistemics and bodies of knowledge can occur. Project funding combining research and innovation activities can help to create such arenas of encounter and exploration temporally.
    - To ensure sustainability, these projects, at best, are embedded into the expertise of the participating institutions. Those usually operate in wider networks of projects extending along the timeline (from past, present into future). Also other cross-sectoral activities as working groups of network organisation (e.g. on new standards) which bring engineering and scientific expertise from the research and implementation front together support this process of co-creation and co-design.
- To which extent new AI driven Information Retrieval approaches as described in our use case empower both content producers and content consumers?
    - The Retrieval-Augmented Generation (RAG) framework is discussed both from a more fundamental methodological angle as well as for various applications in different contributions to these proceedings (see Büttner, 2026; Grasshoff, 2026; Hill, 2026; Meding and Daugs, 2026a; Schlattmann, 2026; and Simons et

> al., 2026). As Meding and Daugs (2026a) formulated "RAG shifts the epistemic burden from the model's internal weights to verifiable information". While RAG by no means is easily traceable and transparent to the user, we agree with Hill (2026) that "RAG occupies an epistemic middle ground, underscoring that researcher judgment and reflexivity are constitutive of knowledge production. So, we see RAG as a promising direction for enhancing the reliability, transparency, and adaptability of AI systems in knowledge-intensive applications. It explicitly builds on existing knowledge as visible in structure information (Merono Penuela et al., 2025). RAG rather aims to empower users to use their own epistemic capacities than to seamlessly (and intransparent) paving their path through the large information spaces.

As these proceedings show, the power of new technologies unfold more forcefully when paired with reflection. If one wishes to influence new technologies, engaging in co-creation from the very start is preferable to debating in the aftermath. The occurrence of lock-in's (Bruckner et al. 1994) seems to be almost unavoidable in technological innovation. In complex systems, as the societies we live in are, it is impossible to find one optimum – right for all needs and interests. But without documenting and reflecting on the processes from a perspective of knowledge and science dynamics – often called meta-level – the chances to influence technological innovation are nil. Such reflections are needed in the sciences themselves, beyond academia adopters, but at best appear as often as possible in those grey areas at the borders between fine and firm lined epistemics spaces where various practices meet. Practices which reach from the creation and preservation of art, performance, culture to engineering machines which support human creativity. The LLM workshop carries the aspiration of such a wide horizon and inspired this specific contribution.[19]

19 In section 1.1. of this chapter DeepL, Grammarly, and ChatGPT have been used for grammar checking and text polishing. Further, no other support of LLM's was used. All model-generated text was reviewed and, where necessary, rewritten by the authors, who remain fully responsible for the final version. For details on the use of LLMs in this volume, see the statement in the volume's introduction.

## Acknowledgement

This paper has been made possible by various research projects. First, the MuseIT project, coordinated by Nasrine Olson at Högskolan i Borås. MuseIT is co-funded by the European Union (Project ID: 101061441). The work presented in this paper also builds on other projects such as Polifonia: a digital harmoniser for musical heritage knowledge (EU funded, Project ID 101004746), and SSHOC-NL, a Dutch funded research infrastructure project in the Social Sciences and Humanities. We would like to thank partners from the Now.Museum initiative (https://www.now.museum/now-museum/) for their support in the exploration. Part of the paper has been informed by participation in the FAIRImpact project (EU funded, Project ID 101057344). Han Yang received funding from the Deutsche Forschungsgemeinschaft (DFG) under grant number: MA 3964/15-3 (SocioHub project). Han Yang and Philipp Mayr received additional funding from the European Union under the Horizon Europe grant OMINO – Overcoming Multilevel INformation Overload under grant number 101086321. We thank Andrew Treloar and CJ Woodford for comments on this paper.